\documentclass[10pt,conference]{IEEEtran}
\usepackage{cite}
\usepackage{amsmath,amsfonts}
\usepackage{dsfont}
\usepackage{graphicx}

\newtheorem{definition}{Definition}
\newtheorem{theorem}{Theorem}
\newtheorem{lemma}{Lemma}

\renewcommand{\vec}[1]{\boldsymbol{#1}}

\newcommand{\cnormal}[1]{\sim\mathcal{CN}(#1)}
\newcommand{\E}[1]{\mathds{E}\left[#1\right]}
\newcommand{\EE}[2]{\mathds{E}_{#1}\left[#2\right]}
\newcommand{\smallE}[1]{\mathds{E}[#1]}
\newcommand{\smallEE}[2]{\mathds{E}_{#1}[#2]}
\newcommand{\prob}[1]{\mathds{P}\left[#1\right]}
\newcommand{\tr}[1]{\text{tr}(#1)}
\newcommand{\e}{\text{e}}
\newcommand{\dpch}[1]{\text{dpch}\big(#1\big)}

\newcommand{\N}{\mathds{N}}
\newcommand{\C}{\mathds{C}}
\newcommand{\R}{\mathds{R}}
\newcommand{\setG}{\mathcal{G}}
\newcommand{\setS}{\mathcal{S}}
\newcommand{\setV}{\mathcal{V}}
\newcommand{\setW}{\mathcal{W}}
\newcommand{\setX}{\mathcal{X}}

\newcommand{\code}{(M_1^{(n)},M_2^{(n)},n)}
\newcommand{\error}[1]{\lambda_{#1}(v)}
\newcommand{\merror}[1]{\mu_{#1}^{(n)}}
\newcommand{\aerror}[1]{\mu_{#1}^{(n)}}

\newcommand{\Rone}{R_{1}}
\newcommand{\Ronestar}{R_{1}^{\star}}
\newcommand{\Roneone}{R_{1}^{(1)}}
\newcommand{\Ronetwo}{R_{1}^{(2)}}
\newcommand{\Rtwo}{R_{2}}
\newcommand{\Rtwostar}{R_{2}^{\star}}
\newcommand{\Rtwoone}{R_{2}^{(1)}}
\newcommand{\Rtwotwo}{R_{2}^{(2)}}
\newcommand{\Rk}{R_{k}}
\newcommand{\Rkstar}{R_{k}^{\star}}

\newcommand{\Pzero}{P_e^{(0)}(v)}
\newcommand{\Pkm}{P_{e,k}^{(m)}(v)}
\newcommand{\Pkone}{P_{e,k}^{(1)}(v)}
\newcommand{\Pktwo}{P_{e,k}^{(2)}(v)}
\newcommand{\Ponetwo}{P_{e,1}^{(2)}(v)}
\newcommand{\Ptwotwo}{P_{e,2}^{(2)}(v)}
\newcommand{\pzero}{P_e^{(0)}}
\newcommand{\pkm}{P_{e,k}^{(m)}}
\newcommand{\pkone}{P_{e,k}^{(1)}}
\newcommand{\pktwo}{P_{e,k}^{(2)}}
\newcommand{\ponetwo}{P_{e,1}^{(2)}}

\newcommand{\eps}[1]{\epsilon_{#1}^{(n)}}

\newcommand{\cregion}{\mathcal{C}_{\text{BDBC}}}
\newcommand{\rregion}{\mathcal{R}_{\text{BDBC}}}

\begin{document}

\title{Capacity of Gaussian MIMO Bidirectional Broadcast Channels}
\author{
  \authorblockN{Rafael F. Wyrembelski\authorrefmark{1},
                Tobias J. Oechtering\authorrefmark{2},
                Igor Bjelakovi\'c\authorrefmark{1},
                Clemens Schnurr\authorrefmark{2}, and
                Holger Boche\authorrefmark{1}\authorrefmark{2}\\[2mm]}
  \authorblockA{\authorrefmark{1}Heinrich-Hertz-Chair for Mobile Communications, Technical University of Berlin\\}
  \authorblockA{\authorrefmark{2}Fraunhofer German-Sino Lab for Mobile Communications\\
                                 Einsteinufer 25/37, D-10587 Berlin, Germany\\[1mm]
                                 Email: \{rafael.wyrembelski, igor.bjelakovic, holger.boche\}@mk.tu-berlin.de, \\
                                 \{tobias.oechtering, clemens.schnurr\}@hhi.fraunhofer.de\\}
}
\maketitle

\begin{abstract}
We consider the broadcast phase of a three-node network, where a relay node establishes a bidirectional communication between two nodes using a spectrally efficient two-phase decode-and-forward protocol. In the first phase the two nodes transmit their messages to the relay node. Then the relay node decodes the messages and broadcasts a re-encoded composition of them in the second phase. We consider Gaussian MIMO channels and determine the capacity region for the second phase which we call the Gaussian MIMO bidirectional broadcast channel.
\end{abstract}

\section{Introduction}
\label{sec:introduction}

Future wireless communication systems should provide high data rates reliably in a certain area, even if the direct link does not have the desired quality due to path loss or shadowing. To face this challenge there has been growing interest in cooperative protocols where some nodes act as relay nodes to guarantee a closed coverage by multi-hop communication. In this work, we consider a three-node network, where one relay node establishes a bidirectional communication between the two other nodes. The problem of the two-way communication without a relay node is first studied in \cite{Sha61twcc}.

\begin{figure}
	\centering
		\scalebox{0.35}{\includegraphics[width=1.00\textwidth]{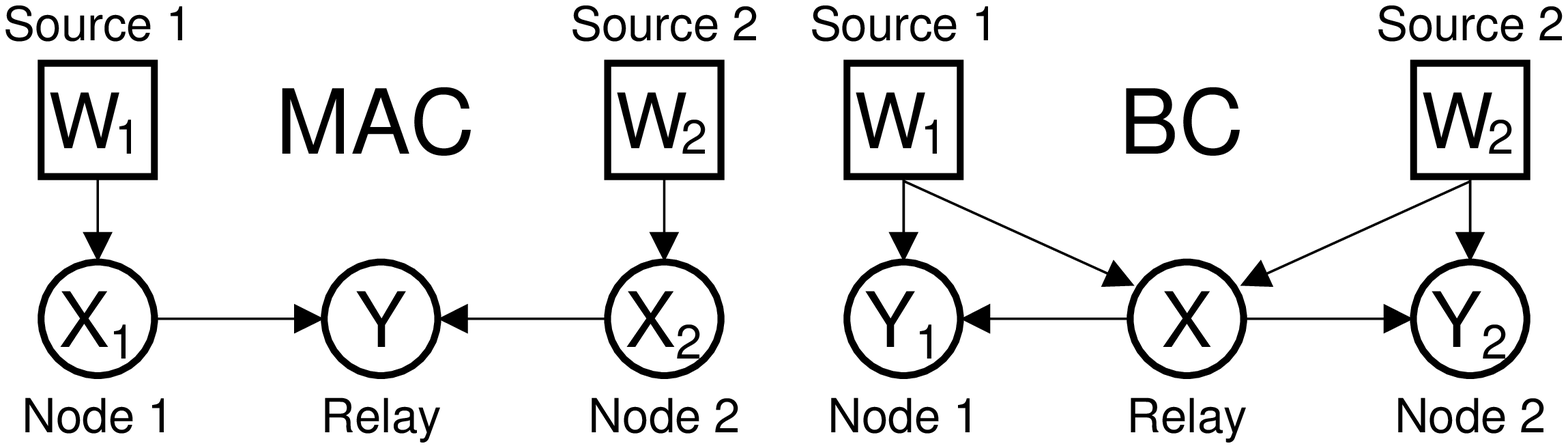}}
	\vspace{-0.3cm}\caption{Multiple access and broadcast phase of the bidirectional relay channel.}
	\label{fig:phases}
\end{figure}

Since it is difficult to isolate simultaneously transmitted and received wireless signals within the same frequency band, we assume half-duplex nodes and therefore allocate orthogonal resources in time for orthogonal transmission and reception. Accordingly, the whole transmission is separated into two phases as depicted in Figure \ref{fig:phases}. In the first phase of a decode-and-forward protocol both nodes transmit their information to the relay node and in the second phase the relay node decodes the messages and broadcasts a re-encoded composition of them. Since we do not allow any cooperation between nodes 1 and 2, this can be seen as a restricted two-way relay channel. 

We assume multiple antennas at all nodes since they can increase the capacity of a system significantly \cite{BCCG07mwc}. The optimal coding strategy for the Gaussian multiple access (MAC) phase is well known \cite{Ahl71mwcc}, \cite{Lia72mac} and extends to the Gaussian MIMO case, see for instance \cite{BCCG07mwc}. We can assume that the relay node can successfully decode the messages $w_1\in\setW_1$ and $w_2\in\setW_2$ from nodes 1 and 2 if we choose the corresponding rate pair within the capacity region. In the following bidirectional broadcast phase $w_1$ is known at the relay node and node 1 and $w_2$ is known at the relay node and node 2. It remains for the relay node to broadcast a message which allows both nodes to recover the unknown message.

The bidirectional broadcast phase is analyzed for the discrete memoryless channel with finite alphabets in \cite{OSBB07bcro}. An achievable rate region of a compress-and-forward approach, where the relay node broadcasts a compressed version of the MAC output to both nodes, can be found in \cite{SOS07arft}. 

In this work, we extend the protocol of \cite{OSBB07bcro} to the Gaussian MIMO bidirectional broadcast channel. The capacity region for this case cannot be given in closed form because of its complicated structure. Therefore we use convex optimization methods to characterize the boundary of the capacity region.\footnote{\emph{Notation}: Matrices and random variables are denoted by bold capital letters, vectors by bold lower case letters, and sets by calligraphic letters; $\R_+$ denotes the set of non-negative real numbers and $\mathbb{M}(N,\C)$ the space of $N\times N$ matrices with complex entries; $(\cdot)^{-1}$ and $(\cdot)^H$ denote inverse and Hermitian transpose; $\E{\cdot}$ is the expectation; $\vec{Q}\succeq0$ means $\vec{Q}$ is positive semidefinite; $\lim\inf$ and $\lim\sup$ denote the limit inferior and limit superior.}

\section{Preliminaries}
\label{sec:preliminaries}

Let $N_R$ be the number of transmit antennas at the relay node and $N_k$ be the number of receive antennas at node $k$, $k=1,2$. We define the discrete-time, memoryless Gaussian MIMO channels between the relay node and nodes 1 and 2 respectively as linear time-invariant multiplicative channels with additive white Gaussian noise. The vector-valued linear input-output relations at one time instant can be expressed as
\begin{equation}
	\vec{y}_k = \vec{H}_k\vec{x} + \vec{n}_k, \quad k=1,2,
	\label{eq:channel}
\end{equation}
where $\vec{y}_k\in\C^{N_k\times1}$ denotes the output, $\vec{H}_k\in\C^{N_k\times N_R}$ the channel matrix, $\vec{x}\in\C^{N_R\times1}$ the input, and $\vec{n}_k\in\C^{N_k\times1}$ the complex circular symmetric distributed Gaussian noise of the channel according to $\mathcal{CN}(\vec{0},\sigma^2\vec{I}_{N_k})$.

We assume that the input alphabet is continuous so that it is reasonable to consider an input constraint. A common and physically meaningful constraint is an average power constraint. This means any sequence $\vec{x}_1,\vec{x}_2,...,\vec{x}_n$ of length $n$ must satisfy
\begin{equation}
  \tfrac{1}{n}\sum_{i=1}^n{\vec{x}_i^H\vec{x}_i}\leq P.
  \label{eq:power}
\end{equation}

\begin{definition}
The \emph{Gaussian MIMO bidirectional broadcast channel with average power limitation} consists of two channels between the relay node and nodes 1 and 2 as defined in (\ref{eq:channel}) with $\vec{x}\in\setX\subset\C^{N_R\times1}$, where $\setX$ describes the set of possible input sequences which satisfy the average power constraint (\ref{eq:power}), i.e., for a sequence of length $n$ we have $\setX^n := \{(\vec{x}_1,\vec{x}_2,...,\vec{x}_n)\in\C^{N_R\times n}:\tfrac{1}{n}\sum_{i=1}^n{\vec{x}_i^H\vec{x}_i}\leq P\}$.
\end{definition}

Let $W_1$ and $W_2$ be the independent information sources at nodes 1 and 2, which are also known at the relay node. We assume that $W_k$ is uniformly distributed on the message set $\setW_k := \{1,2,...,M_k^{(n)}\}$ with $n$ the length of the block code. Further, we use the abbreviation $\setV := \setW_1\times\setW_2$.

\begin{definition}
A \emph{$\code$-code} for the Gaussian MIMO bidirectional broadcast channel with average power limitation consists of one encoder at the relay node
\begin{equation*}
	f: \setV \rightarrow \setX^n
\end{equation*}
and decoders at nodes 1 and 2
\begin{align*}
  g_1: \C^{N_1\times n} \times \setW_1 \rightarrow \setW_2 \cup \{0\}, \\
  g_2: \C^{N_2\times n} \times \setW_2 \rightarrow \setW_1 \cup \{0\}.
\end{align*}
The element $0$ in the decoder plays the role of an erasure symbol and is included in the definition for convenience only.
\end{definition}

When the relay node sends the message $v = [w_1,w_2]$, the receiver of node 1 is in error if $g_1(\vec{y}_1^n,w_1) \neq w_2$.  We denote the probability of this event by $\error{1} := \prob{g_1(\vec{y}_1^n,w_1)\neq w_2 \,|\, f(v) \text{ has been sent}}$. Accordingly, we denote the probability that the receiver of node 2 is in error by $\error{2} := \prob{g_2(\vec{y}_2^n,w_2)\neq w_1 \,|\, f(v) \text{ has been sent}}$. Now, we are able to introduce the notation for the average probability of error for the $k$-th node
\begin{equation*}
	\aerror{k} := \tfrac{1}{|\setV|}\sum_{v\in\setV}{\error{k}}.
\end{equation*}

\begin{definition}
A rate pair $[\Rone,\Rtwo]$ is said to be \emph{achievable} for the Gaussian MIMO bidirectional broadcast channel with average power limitation if for any $\delta>0$ there is an $n(\delta)\in\N$ and a sequence of $\code$-codes satisfying the power constraint such that for all $n \geq n(\delta)$ we have $\frac{\log M_1^{(n)}}{n}\geq\Rtwo-\delta$ and $\frac{\log M_2^{(n)}}{n}\geq\Rone-\delta$ while $\merror{1}, \merror{2}\rightarrow0$ as $n\rightarrow\infty$. The capacity region is the set of all achievable rate pairs which is defined as $\cregion:=\{[R_1,R_2]\in\R_+^2: [R_1,R_2] \text{ achievable}\}$.
\end{definition}

\section{Capacity Region}
\label{sec:region}

In this section we present and prove our main result which is the capacity region of the Gaussian MIMO bidirectional broadcast channel.

\begin{theorem}
For given covariance matrix $\vec{Q}$ with $\tr{\vec{Q}}\leq P$ satisfying the power constraint the corresponding rate pair $[C_1(\vec{Q}),C_2(\vec{Q})]$ is given by
\begin{align*}
  C_k(\vec{Q}) &:= \log\det(\vec{I}_{N_k}+\tfrac{1}{\sigma^2}\vec{H}_k\vec{Q}\vec{H}_k^H), \quad k=1,2.
\end{align*}
Then the capacity region $\cregion$ of the Gaussian MIMO bidirectional broadcast channel is given by
\begin{equation*}
  \cregion := \underset{\vec{Q}:\; \tr{\vec{Q}}\leq P,\, \vec{Q}\succeq0}{\bigcup}\; \dpch{[C_1(\vec{Q}),C_2(\vec{Q})]}
\end{equation*}
where $\dpch{\cdot}$ denotes the downward positive comprehensive hull which is defined for the vector $\vec{x}\in\R_+^2$ by the set $\dpch{\vec{x}}:=\{\vec{y}\in\R_+^2:y_i\leq x_i,i=1,2\}$. 
\end{theorem}

\subsection{Proof of Achievability}
\label{sec:achievability}

We follow \cite{OSBB07bcro} and adapt the random coding proof for the degraded broadcast channel of \cite{Ber73rctf} to our context.

For a given covariance matrix $\vec{Q}$ with $\tr{\vec{Q}}\leq P$ satisfying the power constraint we have to show that all rate pairs $[\Rone,\Rtwo]$ are achievable which satisfy
\begin{align*}
  \Rk &\leq \log\det(\vec{I}_{N_k}+\tfrac{1}{\sigma^2}\vec{H}_k\vec{Q}\vec{H}_k^H), \quad k=1,2.
\end{align*}
We denote the achievable rate region as $\rregion$.

\subsubsection{Random codebook generation and encoding}

For any $\delta>0$ we have to ensure that the probability that a codeword does not satisfy the power constraint goes to zero as the block length $n$ goes to infinity. Therefore, we define the covariance matrix $\vec{\hat{Q}} := \frac{\hat{P}}{P}\vec{Q}$ with $\hat{P} := P-\epsilon_P$, $\epsilon_P\in(0,P]$, where $\epsilon_P$ allows us to get the rate $\hat{R}_k$ corresponding to the transmit strategy $\vec{\hat{Q}}$ arbitrarily close to $\Rk$. Then we define for any $\hat{R}_k$ the rate of the code $\Rkstar:=\max\{\frac{1}{n}\lfloor n(\hat{R}_k-\frac{\delta}{2})\rfloor,0\}$, $k=1,2$.\footnote{If $\Rkstar=0$, the error probability is zero by definition so that we always assume $\Rkstar>0$ in the following.} 

We generate $M_1^{(n)}M_2^{(n)}$ codewords of length $n$ with $M_1^{(n)}:=2^{n\Rtwostar}$ and $M_2^{(n)}:=2^{n\Ronestar}$, where for each $v=[w_1,w_2]\in\setV$ each entry of the corresponding codeword $f(v)=\vec{x}^n(v)$ is independently chosen according to $\mathcal{CN}(\vec{0},\vec{\hat{Q}})$. We first bound the error probability with respect to the codebook which might violate the power constraint. Therefore let $\hat{\lambda}_k(v)$ and $\hat{\mu}_k^{(n)}$, $k=1,2$, be the corresponding error probabilities. 

In the following the random variable $\vec{X}$ denotes an entry of the codeword $\vec{X}^n$ and the random variable $\vec{Y}_{\!k}$ an entry of the output $\vec{Y}_k^n$, $k=1,2$.

\subsubsection{Decoding}

The receiving nodes use typical set decoding. Let $I(\vec{X};\vec{Y}_k):=\EE{\vec{X}^n,\vec{Y}_k^n}{i(\vec{X}^n;\vec{Y}_k^n)}$, $k=1,2$, denote the average mutual information with $i(\vec{x}^n;\vec{y}_k^n):=\frac{1}{n}\log\frac{p(\vec{y}_k^n|\vec{x}^n)}{p(\vec{y}_k^n)}$ for realizations $\vec{x}^n,\vec{y}_k^n$ of the random variables $\vec{X}^n,\vec{Y}_k^n$. At each receiving node $k$ we have the decoding sets
\begin{equation*}
	\setS(\vec{y}_k^n) := \bigg\{ \vec{x}^n\in\setX^n: i(\vec{x}^n;\vec{y}_k^n) > \frac{\Rkstar+I(\vec{X};\vec{Y}_{\!k})}{2} \bigg\}
\end{equation*}
and the indicator function
\begin{align*}
  d(\vec{x}^n,\vec{y}_k^n) := \begin{cases}1, \quad &\text{if } \vec{x}^n \notin \setS(\vec{y}_k^n) \\
                                           0, \quad &\text{else}. \end{cases}
\end{align*}
When $\vec{x}^n$ has been sent, and $\vec{y}_1^n$ and $\vec{y}_2^n$ have been received at nodes 1 and 2, two different events of error may occur at the decoder: the codeword $\vec{x}^n$ is not in $\setS(\vec{y}_k^n)$ (occurring with probability $\Pkone$) and at node one $\vec{x}^n(w_1,\hat{w}_2)$ with $\hat{w}_2\neq w_2$ is in $\setS(\vec{y}_1^n)$ or at node two $\vec{x}^n(\hat{w}_1,w_2)$ with $\hat{w}_1\neq w_1$ is in $\setS(\vec{y}_2^n)$ (occurring with probability $\Pktwo$). If there is no or more than one codeword $\vec{x}^n(w_1,\cdot)\in\setS(\vec{y}_1^n)$ or $\vec{x}^n(\cdot,w_2)\in\setS(\vec{y}_2^n)$, the decoders map on the erasure symbol $0$.

\subsubsection{Analysis of the probability of decoding error}

From the union bound we have $\hat{\lambda}_k(v)\leq\Pkone+\Pktwo$ with 
\begin{align*}
  \Pkone &\!:=\!\int_{\C^n}{p\big(\vec{y}_k^n|\vec{x}^n(v)\big)d\big(\vec{x}^n(v),\vec{y}_k^n\big)\,d\vec{y}_k^n}\;\,\text{for } k=1,2,\\
  \Ponetwo &\!:=\! \int_{\C^n}{\!\!p\big(\vec{y}_1^n|\vec{x}^n(v)\big)\!\!\!\sum_{\substack{\hat{w}_2=1 \\ \hat{w}_2 \neq w_2}}^{|\setW_2|}{\!\!\!\big(1\!-\!d\big(\vec{x}^n(w_1,\hat{w}_2),\vec{y}_1^n\big)\big)} d\vec{y}_1^n}.
\end{align*}
The error event $\Ptwotwo$ is defined similarly. For uniformly distributed messages $W_1$ and $W_2$ we define $\pkm:=\frac{1}{|\setV|}\sum_{v\,\in\,\setV}{\Pkm}$, $m=1,2$, so that $\hat{\mu}_k^{(n)}\leq\pkone+\pktwo$.

For applying the weak law of large numbers we have to ensure that the first two moments are finite \cite[Section 7.3]{Wil91pwm}.
\begin{lemma}
The mean and variance of $i(\vec{X}^n;\vec{Y}_k^n)$, $k=1,2$, are finite.
\begin{proof}
The proof is a generalization of \cite[Theorem 8.2.2]{Ash65it} to the vector-valued case and is omitted here for brevity.
\end{proof}
\end{lemma}

Next, we average over all codebooks and show that $\smallEE{\vec{X}^n}{\pkone}, \smallEE{\vec{X}^n}{\pktwo} \rightarrow0$ as $n\rightarrow\infty$ if $\hat{R}_k\leq I(\vec{X};\vec{Y}_k)$, $k=1,2$. Recall that $\Rkstar\leq\hat{R}_k-\frac{\delta}{2}$ holds so that we have
\begin{align*}
  \smallEE{\vec{X}^n}{\pkone} &= \frac{1}{|\setV|}\sum_{v\in\setV}\smallEE{\vec{X}^n}{\pkone(v)} \\
    &= \EE{\vec{X}^n}{\int_{\C^n}{p\big(\vec{y}_k^n|\vec{X}^n(v)\big)d\big(\vec{X}^n(v),\vec{y}_k^n\big)} \,d\vec{y}_k^n} \\
    &= \int_{\C^n}\int_{\C^n}{p\big(\vec{x}^n\big)p\big(\vec{y}_k^n|\vec{x}^n\big)
       d\big(\vec{x}^n,\vec{y}_k^n\big) \,d\vec{y}_k^n \,d\vec{x}^n} \\
    &= \EE{\vec{X}^n,\vec{Y}_k^n}{d\big(\vec{X}^n,\vec{Y}_k^n\big)} \!=\! \prob{d\big(\vec{X}^n,\vec{Y}_k^n\big)=1} \\
    &= \prob{i(\vec{x}^n;\vec{y}_k^n) \leq \frac{\Rkstar+I(\vec{X};\vec{Y}_{\!k})}{2}} \\
    &\leq \prob{i(\vec{x}^n;\vec{y}_k^n) \leq I(\vec{X};\vec{Y}_{\!k})-\frac{\delta}{4}} \underset{n\rightarrow\infty}{\rightarrow} 0
\end{align*}
by the law of large numbers since Lemma 1 holds. The fourth equality follows from Fubini's theorem. For the calculation of $\smallEE{\vec{X}^n}{\pktwo}$ we have to distinguish between the nodes. We present the analysis for $k=1$, the case $k=2$ follows similarly. We use the fact that for $v=[w_1,w_2]\neq[w_1,\hat{w}_2]$ the random variables in $p(\vec{y}_1^n|\vec{X}^n(v))$ and $d(\vec{X}^n(w_1,\hat{w}_2),\vec{y}_1^n)$ are independent for each choice of $\vec{y}_1^n\in\C^n$.
\begin{align*}
  &\smallEE{\vec{X}^n}{\ponetwo} = \frac{1}{|\setV|}\sum_{v\in\setV}\smallEE{\vec{X}^n}{\ponetwo(v)} \\
    = &\mathds{E}_{\vec{X}^n\!\!}\Bigg[\int_{\C^n}{\!\!p\big(\vec{y}_1^n|\vec{X}^n(v)\big)
        \!\!\!\sum_{\substack{\hat{w}_2=1 \\ \hat{w}_2 \neq w_2}}^{|\setW_2|}
        {\!\!\!\big(1\!-\!d\big(\vec{X}^n(w_1,\hat{w}_2),\vec{y}_1^n\big)\big)}} d\vec{y}_1^n\Bigg] \\
    = &\!\!\int_{\C^n}\!\!\!\sum_{\substack{\hat{w}_2=1 \\ \hat{w}_2 \neq w_2}}^{|\setW_2|}
        \!\!\!\!\EE{\vec{X}^n\!\!\!}{p\big(\vec{y}_1^n|\vec{X}^n\!(v)\big)\!}
        \!\EE{\vec{X}^n\!\!\!}{\!1\!-\!d\big(\vec{X}^n\!(w_1,\!\hat{w}_2),\vec{y}_1^n\big)\!}\!\! d\vec{y}_1^n \\
    = &\!\int_{\C^n}{\sum_{\substack{\hat{w}_2=1 \\ \hat{w}_2 \neq w_2}}^{|\setW_2|}{p\big(\vec{y}_1^n\big)
        \EE{\vec{X}^n}{1-d\big(\vec{X}^n(w_1,\hat{w}_2),\vec{y}_1^n\big)}} \,d\vec{y}_1^n} \\
    = &\!\int_{\C^n}{\!\!\sum_{\substack{\hat{w}_2=1 \\ \hat{w}_2 \neq w_2}}^{|\setW_2|}{\!\!\!p\big(\vec{y}_1^n\big)
        \!\!\int_{\C^n}{\!p\big(\vec{x}^n\big)
        \big(1\!-\!d\big(\vec{x}^n(w_1,\hat{w}_2),\vec{y}_1^n\big)\big) d\vec{x}^n}} d\vec{y}_1^n} \\
    = &(|\setW_2|-1)\int_{\C^n}\int_{\setS(\vec{y}_1^n)}{p\big(\vec{x}^n\big)p\big(\vec{y}_1^n\big) \,d\vec{x}^n \,d\vec{y}_1^n},
\end{align*}
where in the third equality the change of the order of integration follows again from Fubini's theorem. Whenever $\vec{x}^n\in\setS(\vec{y}_1^n)$, we have $i(\vec{x}^n;\vec{y}_1^n) = \frac{1}{n}\log\frac{p(\vec{y}_1^n|\vec{x}^n)}{p(\vec{y}_1^n)} > \frac{1}{2}(\Ronestar+I(\vec{X};\vec{Y}_{\!1}))$ or $p(\vec{y}_1^n)<p(\vec{y}_1^n|\vec{x}^n) 2^{-\frac{n}{2}(\Ronestar+I(\vec{X};\vec{Y}_{\!1}))}$. Consequently,
\begin{align*}
  \smallEE{\vec{X}^n}{\ponetwo} &< |\setW_2|\int_{\C^n}\int_{\setS(\vec{y}_1^n)}
    p\big(\vec{x}^n\big)p\big(\vec{y}_1^n|\vec{x}^n\big) \\
    &\qquad\qquad \times2^{-\frac{n}{2}
    (\Ronestar+I(\vec{X};\vec{Y}_{\!1}))} \,d\vec{x}^n \,d\vec{y}_1^n \\
    &\leq 2^{n\Ronestar}2^{-\frac{n}{2}(\Ronestar+I(\vec{X};\vec{Y}_{\!1}))} = 2^{\frac{n}{2}(\Ronestar-I(\vec{X};\vec{Y}_{\!1}))} \\
    &\leq 2^{\frac{n}{2}(\hat{R}_1-\frac{\delta}{2}-I(\vec{X};\vec{Y}_{\!1}))}
     \leq 2^{-\frac{n\delta}{4}} \underset{n\rightarrow\infty}\rightarrow 0
\end{align*}
if $\hat{R}_1\leq I(\vec{X},\vec{Y}_1)$. The case $k=2$ follows immediately so that if $\hat{R}_k\leq I(\vec{X};\vec{Y}_{\!k})$, $k=1,2$, the average probability of error gets arbitrarily small for sufficiently large block length~$n$.

\subsubsection{Codebook that satisfies the power constraint}

Up to now some codewords $f(v)$ may violate the power constraint. The probability of this event is given by
\begin{align*}
	\Pzero   := \prob{\tfrac{1}{n}\|\vec{X}^n(v)\|^2>P}.
\end{align*}
Next, for each randomly generated codebook we construct a new codebook where we choose for all codewords $f(v)$, which do not satisfy the power constraint, the zero sequence instead which obviously satisfies the power constraint. We upper bound the probability of a decoding error of the zero sequence with $1$. Then it easily follows
\begin{equation*}
	\error{k} \leq \Pzero + \Pkone + \Pktwo, \quad k=1,2.
\end{equation*}
Since $W_1$ and $W_2$ are uniformly distributed, we have $\pzero := \frac{1}{|\setV|}\sum_{v\,\in\,\setV}{\Pzero}$ and $\pkm:=\frac{1}{|\setV|}\sum_{v\,\in\,\setV}{\Pkm}$, $m=1,2$, so that $\mu_k^{(n)} \leq \pzero+\pkone+\pktwo$. Averaging over all codebook realizations, we get
\begin{equation*}
	\smallE{\mu_k^{(n)}} \leq \smallE{\pzero} + \smallE{\pkone} + \smallE{\pktwo}.
\end{equation*}
The first term describes the probability of violating the power constraint. Thereby $\frac{1}{n}\|\vec{X}^n\|^2$ is the arithmetic average of $n$ independent, identically distributed random variables with $\smallE{\|\vec{X}\|^2}=\hat{P}$. By the weak law of large numbers, the arithmetic average converges in probability to $\hat{P}$. Since $\hat{P}<P$, we have $\smallE{\pzero}\rightarrow0$ as $n\rightarrow\infty$. Since $\smallEE{\vec{X}^n}{\pkone}, \smallEE{\vec{X}^n}{\pktwo} \rightarrow0$ as $n\rightarrow\infty$ as well, we have $\mu_k^{(n)}\rightarrow0$ as $n\rightarrow\infty$, $k=1,2$.



\subsubsection{Achievable rates}

Since $\vec{Y}_{ki}=\vec{H}_{k}\vec{X}_i+\vec{N}_{ki}$, $k=1,2,$ and $\vec{X}_i\cnormal{\vec{0},\vec{\hat{Q}}}$ with $\tr{\vec{\hat{Q}}}=\hat{P}$ and $\vec{N}_{ki}\cnormal{\vec{0},\sigma^2\vec{I}_{N_k}}$ are independent and multivariate normal distributed, it follows that $I(\vec{X};\vec{Y}_{\!k})=\log\det(\vec{I}_{N_k}+\tfrac{1}{\sigma^2}\vec{H}_k\vec{\hat{Q}}\vec{H}_k^H)$ with $\tr{\vec{\hat{Q}}}=\hat{P}$. It exists an $\epsilon_P>0$ such that 
\begin{align*}
  &\hat{R}_k = \log\det(\vec{I}_{N_k}+\tfrac{1}{\sigma^2}\vec{H}_k\vec{\hat{Q}}\vec{H}_k^H) \\
  >&\log\det(\vec{I}_{N_k}+\tfrac{1}{\sigma^2}\vec{H}_k\vec{Q}\vec{H}_k^H)	- \tfrac{\delta}{2} = \Rk - \tfrac{\delta}{2}.
\end{align*}
Finally, we have 
\begin{align*}
	  \Rkstar
	> \log\det(\vec{I}_{N_k}+\tfrac{1}{\sigma^2}\vec{H}_k\vec{Q}\vec{H}_k^H)	- \delta = \Rk - \delta
\end{align*}
while $\merror{k}\rightarrow0$ as $n\rightarrow\infty$, $k=1,2,$ which proves the achievability. \endproof

\subsection{Proof of Weak Converse}
\label{sec:converse}

We have to show that any given sequence of $\code$-codes with $\aerror{1},\aerror{2}\rightarrow0$ there exists a covariance matrix $\vec{Q}$ with $\tr{\vec{Q}}\leq P$ satisfying the power constraint such that 
\begin{align*}
R_1 &\!:= \!\underset{n\rightarrow\infty}{\lim\inf}\tfrac{1}{n}\!\log\! M_2^{(n)}\!\!\leq\!\log\det(\vec{I}_{N_1}\!\!+\!\tfrac{1}{\sigma^2}\vec{H}_1\vec{Q}\vec{H}_1^H) \!+\! o(n^0) \\
R_2 &\!:= \!\underset{n\rightarrow\infty}{\lim\inf}\tfrac{1}{n}\!\log\! M_1^{(n)}\!\!\leq\!\log\det(\vec{I}_{N_2}\!\!+\!\tfrac{1}{\sigma^2}\vec{H}_2\vec{Q}\vec{H}_2^H) \!+\! o(n^0)
\end{align*}
are satisfied. That means we have $\cregion\subseteq\rregion$.

\begin{lemma}
\label{lem:fano}
For our context we have Fano's inequality
\begin{equation*}
	H(W_2|\vec{Y}_1^n,W_1) \leq \merror{1}\log M_2^{(n)}+1=n\eps{1}
	\label{eq:fano}
\end{equation*}
with $\eps{1}=\frac{\log M_2^{(n)}}{n}\merror{1}+\frac{1}{n}\rightarrow0$ for $n\rightarrow\infty$ as $\merror{1}\rightarrow0$.
\begin{proof}
From $\vec{Y}_1^n$ and $W_1$ node 1 estimates the index $W_2$ from the sent codeword $\vec{X}^n(W_1,W_2)$. We define the event of an error at node 1 as
\begin{align*}
  E_1 := \begin{cases}1, \quad &\text{if } g_1(\vec{Y}_1^n,W_1)\neq W_2 \\
                      0, \quad &\text{if } g_1(\vec{Y}_1^n,W_1)= W_2 \end{cases}
\end{align*}
so that we have for the average probability of error $\aerror{1}=\prob{E_1=1}$. From the chain rule for entropies \cite[Lemma 8.3.2]{Ash65it} we have
\begin{align*}
  H(E_1,W_2|\vec{Y}_1^n,W_1) &\!=\! H(W_2|\vec{Y}_1^n,W_1) \!+\! H(E_1|\vec{Y}_1^n,W_1,W_2) \\
                             &\!=\! H(E_1|\vec{Y}_1^n,W_1) \!+\! H(W_2|E_1,\vec{Y}_1^n,W_1).
\end{align*}
Since $E_1$ is a function of $W_1$, $W_2$, and $\vec{Y}_1^n$, we have $H(E_1|\vec{Y}_1^n,W_1,W_2)=0$. Further, since $E_1$ is a binary-valued random variable, we get $H(E_1|\vec{Y}_1^n,W_1)\leq H(E_1)\leq 1$. So that finally with the next inequality
\begin{align*}
    &H(W_2|\vec{Y}_1^n,W_1,E_1) \\
  = &\prob{E_1=0}H(W_2|\vec{Y}_1^n,W_1,E_1=0) + \\
    &\qquad\prob{E_1=1}H(W_2|\vec{Y}_1^n,W_1,E_1=1) \\
  \leq &(1-\aerror{1})0 + \aerror{1}\log(M_2^{(n)}-1) \leq \merror{1}\log M_2^{(n)}
\end{align*}
we get Fano's inequality for our context.
\end{proof}
\end{lemma}

With a similar derivation we get $H(W_1|\vec{Y}_2^n,W_2) \leq \merror{2}\log M_1^{(n)}+1=n\eps{2}$ with $\eps{2}=\frac{\log M_1^{(n)}}{n}\merror{2}+\frac{1}{n}\rightarrow0$ for $n\rightarrow\infty$ as $\merror{2}\rightarrow0$.

\begin{lemma}
\label{lem:single}
The rate $\frac{1}{n}H(W_2)$ can be bounded as follows
\begin{equation*}
	\tfrac{1}{n}H(W_2) \leq \log\det\big(\vec{I}_{N_1}+
     \tfrac{1}{\sigma^2}\vec{H}_1\big(\tfrac{1}{n}\sum_{i=1}^n{\vec{Q}_i}\big)\vec{H}_1^H\big) + \eps{1}
\end{equation*}
with $\eps{1}=\frac{\log M_2^{(n)}}{n}\merror{1}+\frac{1}{n}\rightarrow0$ for $n\rightarrow\infty$ as $\merror{1}\rightarrow0$.
\begin{proof}
First, we bound the entropy $H(W_2)$ as follows
\begin{align*}
  H(W_2) &= H(W_2|W_1) 
         = I(W_2;\vec{Y}_1^n|W_1)\!+\!H(W_2|\vec{Y}_1^n,W_1) \\
         &\leq I(W_2;\vec{Y}_1^n|W_1) \!+\! n\eps{1} 
         \leq I(W_1,W_2;\vec{Y}_1^n) \!+\! n\eps{1} \\
         &\leq I(\vec{X}^n;\vec{Y}_1^n) + n\eps{1}
\end{align*}
where the equalities and inequalities follow from the independence of $W_1$ and $W_2$, the definition of mutual information, Lemma \ref{lem:fano}, and the chain rule for mutual information. Since $(W_1,W_2)$, $\vec{X}^n$, $\vec{Y}_1^n$ form a Markov chain, it can be shown that the last inequality holds. If we use the definition of mutual information and the memoryless property of the channel, we can express the inequality as
\begin{align*}
  &H(W_2) \leq \big(h(\vec{Y}_1^n) - h(\vec{Y}_1^n|\vec{X}^n)\big) +n\eps{1} \\
   &\quad\!\!= \!\!\sum_{i=1}^n{\!\!\big(h(\vec{Y}_{\!1i})\!-\!h(\vec{Y}_{\!1i}|\vec{X}_i)\big)} \!+\! n\eps{1} 
         \!= \!\!\sum_{i=1}^n{\!I(\vec{Y}_{\!1i};\vec{X}_i)} \!+\! n\eps{1}\!.
\end{align*}
If we divide the inequality by $n$ and use again the definition of mutual information we get
\begin{align*}
  \tfrac{1}{n}H(W_2) &\leq 
   \tfrac{1}{n}\sum_{i=1}^n{\big(h(\vec{Y}_{\!1i})-h(\vec{N}_{1i})\big)} + \eps{1}
\end{align*}
with $\vec{Y}_{\!1i}=\vec{H}_1\vec{X}_i+\vec{N}_{1i}$. The random variables $\vec{X}_i$ and $\vec{N}_{1i}$ with $h(\vec{N}_{1i}) =\log\det(\pi\e\,\sigma^2\vec{I}_{N_1})$ are independent. It follows from the entropy maximization theorem that $h(\vec{Y}_{\!1i}) \leq \log\det\big(\pi\e(\sigma^2\vec{I}_{N_1}+\vec{H}_1\vec{Q}_i\vec{H}_1^H)\big)$ with equality if we have Gaussian input, i.e., $\vec{X}_i\cnormal{\vec{0},\vec{Q}_i}$. Therewith we have $h(\vec{Y}_{\!1i})-h(\vec{N}_{1i}) \leq \log\det\big(\vec{I}_{N_1}+\tfrac{1}{\sigma^2}\vec{H}_1\vec{Q}_i\vec{H}_1^H\big)$.
Finally, we get
\begin{align*}
  \tfrac{1}{n}H(W_2) &\leq \tfrac{1}{n}\sum_{i=1}^n{\log\det\big(\vec{I}_{N_1} 
                               + \tfrac{1}{\sigma^2}\vec{H}_1\vec{Q}_i\vec{H}_1^H\big)} + \eps{1} \\
                    &\leq \log\det\big(\vec{I}_{N_1}
                               + \tfrac{1}{\sigma^2}\vec{H}_1\big(\tfrac{1}{n}\sum_{i=1}^n{\vec{Q}_i}\big)\vec{H}_1^H\big) + \eps{1}
\end{align*}
where the second inequality follows from the concavity of the $\log\det$ function \cite[Theorem 7.6.7]{HJ99ma}.
\end{proof}
\end{lemma}

With a similar derivation we get $\frac{1}{n}H(W_1) \leq \log\det\big(\vec{I}_{N_2} + \tfrac{1}{\sigma^2}\vec{H}_2\big(\tfrac{1}{n}\sum_{i=1}^n{\vec{Q}_i}\big)\vec{H}_2^H\big) + \eps{2}$ with $\eps{2}=\frac{\log M_1^{(n)}}{n}\merror{2}+\frac{1}{n}\rightarrow0$ for $n\rightarrow\infty$ as $\merror{2}\rightarrow0$.

It is clear that $R_1 = \lim\inf_{n\rightarrow\infty}\tfrac{1}{n}\log M_2^{(n)} \leq \lim\sup_{n\rightarrow\infty} \tfrac{1}{n}\log M_2^{(n)}$ holds. Since $W_2$ is uniformly distributed, we have $\tfrac{1}{n}\log M_2^{(n)}=\tfrac{1}{n}H(W_2)$ and obtain with Lemma \ref{lem:single}
\begin{equation}
	R_1 \!\leq\! \underset{n\rightarrow\infty}{\lim\sup}\!\Big[\!
	  \log\det\!\big(\vec{I}_{N_1} \!\!+\! \tfrac{1}{\sigma^2}\vec{H}_1\big(\tfrac{1}{n}\!\sum_{i=1}^{n}{\!\vec{Q}_i}\big)\vec{H}_1^H\big) \!+ \epsilon_1^{(n)}\!\Big].
	\label{eq:limsup1}
\end{equation}
Next, we define the compact set $\setG:=\{\vec{Q}\in\mathbb{M}(N_R,\C):\tr{\vec{Q}}\leq P,\vec{Q}\succeq0\}$ with $\tfrac{1}{n}\sum_{i=1}^{n}{\vec{Q}_i}\in\setG$ since $\tfrac{1}{n}\sum_{i=1}^{n}{\vec{Q}_i}\succeq0$ and $\tfrac{1}{n}\sum_{i=1}^{n}{\tr{\vec{Q}_i}}=\tr{\tfrac{1}{n}\sum_{i=1}^{n}{\vec{Q}_i}}\leq P$ hold. This implies that there exists a subsequence $(n_l)_{l\in\N}$ such that $\tfrac{1}{n_l}\sum_{i=1}^{n_l}{\vec{Q}_i}\rightarrow\vec{Q}$ as $n_l\rightarrow\infty$ with $\vec{Q}\in\setG$. Therewith and with the continuity of the $\log\det$ we have
\begin{align}
&\underset{n_l\rightarrow\infty}{\lim\sup}\bigg[
	  \log\det\big(\vec{I}_{N_1} \!+\! \tfrac{1}{\sigma^2}\vec{H}_1\big(\tfrac{1}{n_l}\sum_{i=1}^{n_l}{\vec{Q}_i}\big)\vec{H}_1^H\big) \!+\! \epsilon_1^{(n_l)}\bigg] \nonumber \\
	= &\log\det\big(\vec{I}_{N_1} + \tfrac{1}{\sigma^2}\vec{H}_1\vec{Q}\vec{H}_1^H\big).
	\label{eq:limsup2}
\end{align}
Combining (\ref{eq:limsup1}) and (\ref{eq:limsup2}) we obtain $R_1\leq \log\det\big(\vec{I}_{N_1} + \tfrac{1}{\sigma^2}\vec{H}_1\vec{Q}\vec{H}_1^H\big)$. Using the same subsequence $(n_l)_{l\in\N}$ and arguments we get $R_2\leq \log\det\big(\vec{I}_{N_2} + \tfrac{1}{\sigma^2}\vec{H}_2\vec{Q}\vec{H}_2^H\big)$ which proves the converse. \endproof

\section{Discussion}
\label{sec:discussion}

Since the capacity region is convex, we can completely characterize $\cregion$ by its boundary which corresponds to the weighted rate sum optimal rate pairs. Therefore, we introduce a weight vector $\vec{q}=[q_1,q_2]\in\R_+^2\backslash\{\vec{0}\}$ and express the weighted rate sum for given $\vec{q}$ as $R_{\Sigma}(\vec{Q}) = q_1\Rone(\vec{Q}) + q_2\Rtwo(\vec{Q})$ with $\Rone(\vec{Q}) := \log\det(\vec{I}_{N_1}+\frac{1}{\sigma^2}\vec{H}_1\vec{Q}\vec{H}_1^H)$ and $\Rtwo(\vec{Q}) := \log\det(\vec{I}_{N_2}+\frac{1}{\sigma^2}\vec{H}_2\vec{Q}\vec{H}_2^H)$. The aim is now to find the optimal covariance matrix $\vec{Q}^*(\vec{q})$ with $\tr{\vec{Q}^*(\vec{q})}\leq P$ satisfying the power constraint which maximizes the weighted rate sum. 
This can be expressed as the following optimization problem
\begin{align}
  \underset{\vec{Q}}{\max} &\quad q_1\Rone(\vec{Q}) + q_2\Rtwo(\vec{Q}) \quad
  \text{s.t. }  \tr{\vec{Q}}\leq P, \; \vec{Q}\succeq0.
  \label{eq:sumrateoptimal}
\end{align}
The Lagrangian for this optimization problem is given by
\begin{align*}
  L(\vec{Q},\mu,\!\vec{\Psi}) \!=\! 
    -q_1\Rone\!(\vec{Q}) \!-\! q_2\Rtwo(\vec{Q}) \!-\! \mu\big(P\!-\!\tr{\vec{Q}}\big) \!-\! \tr{\vec{Q\Psi}}.
\end{align*}
Therewith, the covariance matrix maximizing (\ref{eq:sumrateoptimal}) for given $\vec{q}$ is uniquely characterized by the Karush-Kuhn-Tucker conditions
\begin{subequations}
\begin{align}
\label{eq:kkt}
  -\mu\vec{I}_{N_R} + \vec{\Psi} &= -q_1\vec{H}_1^H(\sigma^2\vec{I}_{N_1}+\vec{H}_1\vec{Q}\vec{H}_1^H)^{-1}\vec{H}_1 \nonumber\\
                                &\phantom{=\;\,} -\!q_2\vec{H}_2^H(\sigma^2\vec{I}_{N_2}+\vec{H}_2\vec{Q}\vec{H}_2^H)^{-1}\vec{H}_2, \\
  \vec{Q} &\succeq 0, \quad P \geq \tr{\vec{Q}}, \label{eq:kkt_primal} \\
  \vec{\Psi} &\succeq 0, \quad \mu\geq 0, \label{eq:kkt_dual} \\
  \tr{\vec{Q\Psi}} &= 0, \quad \mu\big(P-\tr{\vec{Q}}\big) = 0, \label{eq:kkt_comp}
\end{align}
\end{subequations}
with primal, dual, and complementary slackness conditions (\ref{eq:kkt_primal}), (\ref{eq:kkt_dual}), and (\ref{eq:kkt_comp}) respectively. 

The weighted rate sum optimal rate pairs describe the curved section of the boundary and are uniquely characterized by (\ref{eq:kkt})-(\ref{eq:kkt_comp}). The two endpoints correspond to the cases, where $\vec{q}$ is chosen to optimize one unidirectional rate. More precisely, the case $q_1>0,q_2=0$ means we want to maximize the rate $\Rone$, for which we can achieve the single-user capacity $\Roneone := \max \Rone(\vec{Q}) = \log\det(\vec{I}_{N_1}+\tfrac{1}{\sigma^2}\vec{H}_1\vec{Q}^{(1)}\vec{H}_1^H)$, where the superscript $^{(1)}$ indicates that the weight vector $\vec{q}$ is chosen to optimize the rate $\Rone$. For $\Rtwo$ this leads to an achievable rate $\Rtwoone = \log\det(\vec{I}_{N_2}+\tfrac{1}{\sigma^2}\vec{H}_2\vec{Q}^{(1)}\vec{H}_2^H)$. Similar, for $q_1=0,q_2>0$ we can achieve $\Rtwotwo = \log\det(\vec{I}_{N_2}+\tfrac{1}{\sigma^2}\vec{H}_2\vec{Q}^{(2)}\vec{H}_2^H)$ and $\Ronetwo = \log\det(\vec{I}_{N_1}+\tfrac{1}{\sigma^2}\vec{H}_1\vec{Q}^{(2)}\vec{H}_1^H)$. 

\begin{figure}
\begin{minipage}{.48\linewidth}
  \centering
  \includegraphics[width=5cm]{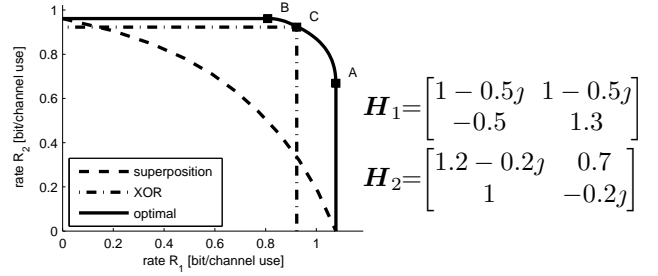}
\end{minipage}
\begin{minipage}{.52\linewidth}
	\centering
	\begin{align*}
		&\vec{H}_1\!\!=\!\!\begin{bmatrix} 1-0.5\jmath & \!\!1-0.5\jmath \\ -0.5 & \!\!1.3\end{bmatrix} \\
		&\vec{H}_2\!\!=\!\!\begin{bmatrix} 1.2-0.2\jmath & \!\!0.7 \\ 1 & \!\!-0.2\jmath\end{bmatrix}
	\end{align*}
\end{minipage}
\caption{Achievable rate regions for $N_1=N_2=N_R=2$.}
\label{fig:bdbc}
\end{figure}

Figure~\ref{fig:bdbc} exemplarily depicts the capacity region in comparison to the achievable rate regions of the superposition coding \cite{OB07otsi} and the XOR coding approach \cite{HKEZ07mtwr}, where the optimal rate pair is given by 
\begin{equation*}
	R_1^* = R_2^* := \max\min\{\Rone(\vec{Q}),\Rtwo(\vec{Q})\}.
\end{equation*}

The optimal unidirectional rate pairs correspond to the points $A$ and $B$ in Figure \ref{fig:bdbc}. Point $C$ in Figure \ref{fig:bdbc} describes the maxmin optimal rate pair, which is the only rate pair where the XOR coding approach achieves the capacity.

\section{Conclusion}
\label{sec:conclusion}

In this work, we extended the bidirectional broadcast phase for the discrete memoryless channel with finite alphabets of \cite{OSBB07bcro} to the Gaussian MIMO case and derived the capacity region. We showed that there is not a unique rate sum optimal transmit strategy. Similar to the Gaussian MIMO MAC the weighted rate sum optimal transmit strategy for the bidirectional broadcast phase depends now on the weights of the two rates.

\bibliographystyle{IEEEtran}
\bibliography{isit_bib}

\begin{thebibliography}{10}
\providecommand{\url}[1]{#1}
\csname url@samestyle\endcsname
\providecommand{\newblock}{\relax}
\providecommand{\bibinfo}[2]{#2}
\providecommand{\BIBentrySTDinterwordspacing}{\spaceskip=0pt\relax}
\providecommand{\BIBentryALTinterwordstretchfactor}{4}
\providecommand{\BIBentryALTinterwordspacing}{\spaceskip=\fontdimen2\font plus
\BIBentryALTinterwordstretchfactor\fontdimen3\font minus
  \fontdimen4\font\relax}
\providecommand{\BIBforeignlanguage}[2]{{%
\expandafter\ifx\csname l@#1\endcsname\relax
\typeout{** WARNING: IEEEtran.bst: No hyphenation pattern has been}%
\typeout{** loaded for the language `#1'. Using the pattern for}%
\typeout{** the default language instead.}%
\else
\language=\csname l@#1\endcsname
\fi
#2}}
\providecommand{\BIBdecl}{\relax}
\BIBdecl

\bibitem{Sha61twcc}
C.~E. Shannon, ``{Two-Way Communication Channel},'' \emph{Proc. 4th Berkeley
  Symp. Math Stat. and Prob.}, vol.~1, pp. 611--644, 1961.

\bibitem{BCCG07mwc}
E.~Biglieri, R.~Calderbank, A.~Constantinides, A.~Goldsmith, A.~Paulraj, and
  H.~V. Poor, \emph{MIMO Wireless Communications}.\hskip 1em plus 0.5em minus
  0.4em\relax Cambr. Univ. Press, 2007.

\bibitem{Ahl71mwcc}
R.~Ahlswede, ``{Multi-Way Communication Channels},'' in \emph{2nd Int. Symp. on
  Inf. Theory}, Tsahkadsor, Armenian, USSR, Sep. 1971, pp. 23--52.

\bibitem{Lia72mac}
H.~Liao, ``{Multiple Access Channels},'' Ph.D. dissertation, University of
  Hawaii, Honolulu, HI, 1972.

\bibitem{OSBB07bcro}
T.~J. Oechtering, C.~Schnurr, I.~Bjelakovic, and H.~Boche, ``{Broadcast
  Capacity Region of Two-Phase Bidirectional Relaying},'' \emph{IEEE
  Transactions on Information Theory}, vol.~54, no.~1, pp. 454--458, Jan. 2008.

\bibitem{SOS07arft}
C.~Schnurr, T.~J. Oechtering, and S.~Stanczak, ``{Achievable Rates for the
  Restricted Half-Duplex Two-Way Relay Channel},'' in \emph{Proc. of the 41st
  Asilomar Conf. on Signals, Systems, and Computers}, Nov. 2007.

\bibitem{Ber73rctf}
P.~P. Bergmans, ``{Random Coding Theorem for Broadcast Channels With Degraded
  Components},'' \emph{IEEE Trans. Inf. Theory}, vol.~19, no.~2, pp. 197--207,
  Mar. 1973.

\bibitem{Wil91pwm}
D.~Williams, \emph{Probability with Martingales}.\hskip 1em plus 0.5em minus
  0.4em\relax Cambr. Univ. Press, 1991.

\bibitem{Ash65it}
R.~B. Ash, \emph{Information Theory}.\hskip 1em plus 0.5em minus 0.4em\relax
  Intersc. Publ., New York, 1965.

\bibitem{HJ99ma}
R.~A. Horn and C.~R. Johnson, \emph{Matrix Analysis}.\hskip 1em plus 0.5em
  minus 0.4em\relax Cambr. Univ. Press, 1999.

\bibitem{OB07otsi}
T.~J. Oechtering and H.~Boche, ``{Optimal Transmit Strategies in Multi-Antenna
  Bidirectional Relaying},'' in \emph{Proc. IEEE Int. Conf. on Acoustics,
  Speech, and Signal Proc.}, Honolulu, HI, Apr. 2007, pp. 145--148.

\bibitem{HKEZ07mtwr}
I.~Hammerstr\"om, M.~Kuhn, C.~Esli, J.~Zhao, A.~Wittneben, and G.~Bauch,
  ``{MIMO Two-Way Relaying with Transmit CSI at the Relay},'' in \emph{Proc.
  IEEE Signal Proc. Adv. in Wireless Comm.}, Helsinki, Finland, Jun. 2007.

\end{thebibliography}

\end{document}